\documentclass[10pt,aps,pra,twocolumn,preprintnumbers,nofootinbib,amsmath,amssymb,superscriptaddress]{revtex4-2}
\usepackage{mathtools}
\usepackage[T1]{fontenc}
\usepackage{amsfonts}
\usepackage{amssymb}
\usepackage{subfiles}
\usepackage{natbib}
\usepackage[colorlinks,linkcolor=red,citecolor=blue,breaklinks=true]{hyperref}
\usepackage{graphicx}
\usepackage{dcolumn}
\usepackage{bm}
\usepackage{color}
\usepackage{amsthm}
\usepackage{dsfont}
\usepackage{tcolorbox}
\usepackage{cases}
\usepackage{physics}

\usepackage{braket}
\usepackage{bbold}
\usepackage{xcolor}
\usepackage[normalem]{ulem}
\usepackage{orcidlink}
\usepackage{lipsum}
\usepackage{tabularx}
\usepackage{booktabs}

\begin{document}

\title{Benchmark of the Schmiedl-Seifert optimal protocol with reduced dissipated work and fluctuations in optical tweezers}

\author{Thalyta T. Martins}
\email{thalyta.tavares.martins@gmail.com}
\affiliation{Instituto de Física de São Carlos, Universidade de São Paulo, 13560-970 São Carlos, SP, Brazil}

\author{André H. A. Malavazi}
\affiliation{International Centre for Theory of Quantum Technologies, University of Gdańsk, ul. prof. Marii Janion 4, 80-309 Gdańsk, Poland
}

\author{Lucas P. Kamizaki}
\affiliation{Instituto de Física de São Carlos, Universidade de São Paulo, 13560-970 São Carlos, SP, Brazil}
\affiliation{Instituto de Física ‘Gleb Wataghin’, Universidade Estadual de Campinas, 13083-859 Campinas, SP, Brazil}

\author{Artyom Petrosyan}
\affiliation{Université de Lyon, ENS de Lyon, CNRS, Laboratoire de Physique, F-69342 Lyon, France 
}

\author{Benjamin Besga}
\affiliation{Université Claude Bernard Lyon 1, CNRS, Institut Lumière Matière, UMR 5306, F69100 Villeurbanne, France
}

\author{Sergio Ciliberto}
\affiliation{Université de Lyon, ENS de Lyon, CNRS, Laboratoire de Physique, F-69342 Lyon, France 
}

\author{ Sérgio R. Muniz}
\email{srmuniz@ifsc.usp.br}
\affiliation{Instituto de Física de São Carlos, Universidade de São Paulo, 13560-970 São Carlos, SP, Brazil}

\date{\today}
             
\begin{abstract}

Finite-time thermodynamic transformations generally dissipate energy, as required by the Second Law, so identifying and investigating energetically optimal processes remain relevant for understanding and designing efficient thermal devices. In this context, we present here a systematic experimental benchmark of the finite-time minimum-work compression protocol derived by Schmiedl and Seifert for an overdamped Brownian particle in a harmonic trap. By parametrically controlling the trap stiffness, we compare the full work statistics of the analytical optimal protocol with those obtained under linear driving across a broad range of protocol durations and amplitudes. We find that the optimal protocol consistently reduces both the mean work and the work variance relative to linear driving. Although the Schmiedl-Seifert protocol was primarily derived to minimize the mean work, our data show, in the present setting, a clear reduction in fluctuations and high-work tails under realistic bandwidth-limited actuation. This observation connects our benchmark to the growing interest in higher-order cumulants and to current developments in stochastic-thermodynamic control.

\end{abstract}

%##########################################################################

\maketitle

%##########################################################################

\section{Introduction} \label{sec:intro}

In classical thermodynamics, the work necessary to perform a process between two equilibrium states of a system in contact with a thermal reservoir is lower bounded by the Helmholtz free-energy difference $\Delta F$~\cite{Callen:450289}. This lower bound represents the reversible limit of the transformation and is attained only in quasi-static isothermal processes, for which the duration of the transformation diverges. 
However, in realistic scenarios, thermodynamic tasks must be performed in finite time, which leads to irreversibility and excess dissipation. This limitation is particularly evident in the well-known trade-off between power and efficiency in finite-time heat engines~\cite{PhysRevLett.95.190602, PhysRevLett.105.150603, PhysRevLett.108.210602, Holubec_2016, PhysRevE.87.012133, Bonança_2019}. 
Interestingly, within the framework of stochastic thermodynamics~\cite{sekimoto1998langevin, Seifert_2012, schuster2013nonequilibrium}, this bound remains valid for microscopic systems driven out of equilibrium when considering the average work over an ensemble of trajectories~\cite{jarzynski1997nonequilibrium}. 

From this perspective, establishing control strategies that minimize dissipation during finite-time transformations is, therefore, of practical relevance and constitutes one of the central goals of optimal control in thermodynamics~\cite{Deffner_2020, Blaber_2023}. Identifying such optimal protocols is crucial for the design of highly efficient finite-time heat engines~\cite{Dechant_2017, martinez2017colloidal} and for improving the accuracy and efficiency of free-energy estimation methods~\cite{PhysRevLett.100.190601, PhysRevE.103.032146}. That said, in small systems, it is rarely enough to discuss the mean cost alone, as fluctuations can be just as consequential.

Most of the optimal-control literature targets the average work or the entropy production~\cite{Deffner_2020, Blaber_2023}. The full spread of work values is nevertheless not just a secondary detail, since large fluctuations are precisely what make nonequilibrium free-energy estimates hard to converge. In the slow, near-equilibrium regime, fluctuations can be tied to dissipation through the Gaussian-cumulant limit of Jarzynski’s equality~\cite{jarzynski1997nonequilibrium}, yielding $\langle W_{\mathrm{diss}}\rangle \equiv \langle W \rangle-\Delta F =\beta\sigma_W^2/2$, that is: a mean dissipated work directly proportional to the work variance. However, in more general cases outside this Gaussian limit, higher-order cumulants can no longer be neglected, and protocols that minimize mean work do not necessarily minimize work variance; they may even differ qualitatively at intermediate-to-fast driving speeds~\cite{Solon-Horowitz2018, Blaber_2023}.

Although finding analytic solutions for optimal protocols can be challenging in the general case, requiring involved mathematical and numerical tools~\cite{Deffner_2020, Blaber_2023}, solutions can be obtained for particular systems~\cite{schmiedl2007optimal,gomez2008optimal, schmiedl2009optimal, Abiuso_2022} and limiting situations, such as the fast-protocol limit~\cite{PhysRevE.104.L022101} and weak processes in the linear response regime~\cite{bonancca2014optimal, PhysRevE.94.052106, bonancca2018minimal, large2019optimal, kamizaki2022performance, Naze_2024}. In this respect, methods for finite-time isothermal transitions, also known as shortcuts to isothermality, have been proposed \cite{PhysRevE.106.054108, PhysRevLett.128.230603, PhysRevE.96.012144} and analyzed experimentally ~\cite{Martinez2016,10.1063/1.4962825, 10.1063/1.5143602, PhysRevResearch.1.033122}. 

Building on these advances, Brownian particles confined in optical potentials constitute a paradigmatic platform for studies in stochastic thermodynamics. The high degree of control over optical forces allows one to precisely tailor the potential landscape experienced by the particle, enabling the implementation of well-defined thermodynamic protocols at the microscopic scale. This level of control has made optical tweezers a powerful and widely used experimental tool for investigating nonequilibrium thermodynamic processes \cite{martins2025brief, ritort2008nonequilibrium}.

In 2007, Schmiedl and Seifert~\cite{schmiedl2007optimal} theoretically derived an analytical optimal protocol for overdamped Brownian particles trapped in harmonic potentials, which provides an excellent representation of a colloidal particle trapped in optical-tweezers. These protocols minimize the average work required to drive the particle between two states in finite time under a time-dependent harmonic potential. 

Recent experiments have demonstrated finite-time equilibration shortcuts, variationally optimized transfers, and optimal-transport protocols in trapped Brownian systems \cite{PhysRevResearch.2.012012,baldovin2025optimal}. The present work addresses a more specific and fundamental benchmark: a direct experimental implementation of the exact minimum-work finite-time Schmiedl-Seifert compression protocol for the overdamped harmonic trap \cite{schmiedl2007optimal}. We do this by comparing the analytical optimum with linear (sub-optimal) driving over a wide range of protocol durations and compression amplitudes, and we show that the optimized protocol reduces both dissipated work and work fluctuations under realistic optical-tweezer constraints.

The remainder of this article is structured as follows. In Sec.~\ref{sec:theo} we briefly review the basic notions of stochastic thermodynamics and present the optimal protocol studied. In Sec.~\ref{sec:exp} we introduce our experimental setup. The results are presented in Sec.~\ref{sec:results}. Finally, in Sec.~\ref{sec:discussion}, we summarize our conclusions and discuss perspectives for future research.  

%##########################################################################

\section{Stochastic thermodynamics and optimal protocols}\label{sec:theo}

A Brownian particle immersed in a thermal bath at temperature $T$ and trapped in a one-dimensional harmonic potential $U(x,\kappa) = \frac{1}{2} \kappa x^2$, with trap stiffness $\kappa$, has its stochastic trajectory described by the overdamped Langevin equation~\cite{jones2015optical}:
\begin{equation}\label{Langevin}
   \gamma \dot{x}  = -\frac{d U(x,\kappa)}{d x} + \mathcal{F}_{th} = -\kappa x + \mathcal{F}_{th},
\end{equation}
where $\gamma=6\pi \eta a$ is the friction coefficient, with $\eta$ being the medium viscosity and $a$ the particle radius. The term $\mathcal{F}_{th}$ is a stochastic force encompassing the thermal fluctuations and is represented as Gaussian white noise, satisfying $\langle \mathcal{F}_{th} (t)\rangle = 0$ and $\langle \mathcal{F}_{th} (t)\mathcal{F}_{th} (t')\rangle = 2\gamma k_B T \delta (t-t')$, where $k_{B}$ is the Boltzmann constant. 

In the compressing protocol, the trapping potential is parametrically changed from a less confining state at equilibrium to a more confining configuration over a time interval $\tau_P$. The protocol is therefore defined by a time-dependent control parameter $\{\lambda(t),\, t \in [0, \tau_P]\}$, with $\lambda(0) = \lambda_i$ and $\lambda(\tau_P) = \lambda_f > \lambda_i$, corresponding to the stiffness of the trap, i.e., $\lambda \equiv \kappa$. 

Such a process is accompanied by energetic exchanges with the thermal bath. Moreover, the stochastic work performed along a single trajectory can be directly computed from the particle's position using the framework introduced by Sekimoto~\cite{sekimoto1998langevin} as
\begin{equation}  W=\int_{\lambda_{i}}^{\lambda_{f}}d\lambda\frac{\partial U(x,\lambda)}{\partial\lambda}=\int_{0}^{\tau_{P}}dt\frac{d\lambda(t)}{dt}\frac{x^{2}(t)}{2}.
    \label{eq:work}
\end{equation}
Analogously, given an ensemble of realizations of the same protocol, the average work reads
\begin{equation}
\langle W\rangle=\int_{\lambda_{i}}^{\lambda_{f}}d\lambda\left\langle \frac{\partial U(x,\lambda)}{\partial\lambda}\right\rangle =\int_{0}^{\tau_{P}}dt\frac{d\lambda(t)}{dt}\frac{\langle x^{2}(t)\rangle}{2}.
\label{eq:avworktheo}
\end{equation}

For instance, in the limit of a very short protocol, $\tau_P \to 0$, the particle does not have time to adjust its position, and the work is determined by the instantaneous change of the trap stiffness:
\begin{equation}
\langle W \rangle_{\tau_P \to 0} =  \frac{k_B T}{2} \left( \frac{\lambda_f}{\lambda_i} - 1 \right).
\label{eq:wstep}
\end{equation} 

Conversely, in the limit of an infinitely slow protocol, $\tau_P \to \infty$, the process approaches a quasi-static evolution, and the average work converges to
\begin{equation}\label{eq:deltaftheo}
\langle W\rangle_{\tau_P \to \infty} = \Delta F \coloneqq F_f - F_i = - k_B T \ln{\sqrt{\frac{\lambda_i}{\lambda_f}}},
\end{equation}
where $F_{i/f} = - k_B T \ln Z_{i/f}$ is the Helmholtz free energy and $Z_{i/f}$ is the corresponding partition function.

For the single-parameter protocols considered here, a finite duration $\tau_P$ generally leads to dissipation. Importantly, the magnitude of this excess work depends on how the control parameter $\lambda(t)$ is varied in time.

\begin{figure}
\centerline{\includegraphics[width=7.0cm]{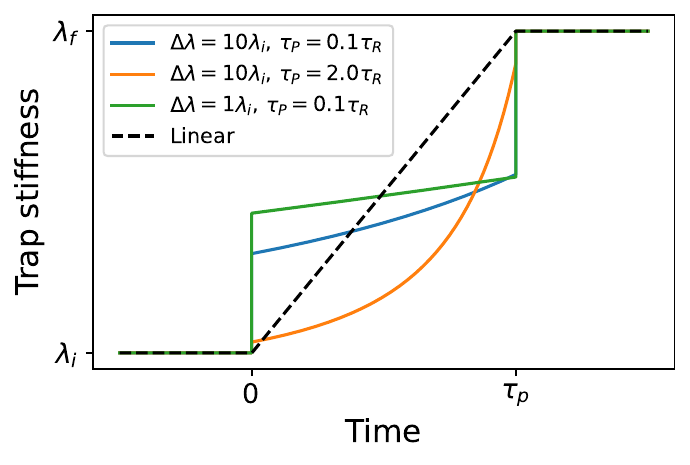}}
\caption{Linear (black dashed line) and optimal (solid lines) protocols obtained for different modulation amplitudes, $\Delta \lambda$, and protocol times, $\tau_P$.}
\label{fig:theoprot}
\end{figure}

One simple and widely used way to vary the control parameter is to change the trap stiffness linearly in time, as shown in Fig. \ref{fig:theoprot} (black dashed line),
\begin{equation} 
\lambda(t) = \lambda_i + \frac{\Delta \lambda}{\tau_P} t, \quad \text{with } \Delta \lambda = \lambda_f - \lambda_i.
\label{eq:linearprotU}
\end{equation}
In this case, the time derivative is constant, $d\lambda/dt = \Delta \lambda / \tau_P$, allowing a straightforward estimation of the average work using Eq.~\eqref{eq:avworktheo} and the time-dependent variance $\langle x^2(t) \rangle$, obtained directly from the Langevin equation\footnote{In the overdamped regime, for each small time step $\Delta t$, the variance evolves as
$$ 
\langle x^2(t+\Delta t) \rangle = \langle x^2(t) \rangle \, e^{-2 \lambda(t) \Delta t / \gamma} + \frac{k_B T}{\lambda(t)} \left( 1 - e^{-2 \lambda(t) \Delta t / \gamma} \right).
$$}.  

However, while simple, this protocol is not the most efficient in terms of minimizing dissipation. In general, each system admits its own optimal driving strategy, and identifying such protocols can be a nontrivial task. For the case of a harmonically trapped Brownian particle undergoing compression in the overdamped limit, Schmiedl and Seifert~\cite{schmiedl2007optimal} showed that the optimal finite-time protocol, $\lambda^*(t)$, is given by\footnote{This result was obtained by solving the Euler-Lagrange equation relative to the work functional Eq.~\ref{eq:avworktheo}.} 
\begin{equation}
    \lambda^*(t)=\frac{\lambda_i-c_2^*\gamma(1+c_2^*t)}{(1+c_2^*t)^2},
    \label{eq:optimalequation}
\end{equation}
where
\begin{equation}
    c_2^*=\frac{-\gamma-\lambda_f\tau_P+\sqrt{\gamma^2+2\lambda_i\tau_P\gamma+\lambda_f\lambda_i\tau_P^2}}{\tau_P(2\gamma+\lambda_f\tau_P)},
\end{equation}
and minimal average work is
\begin{equation}
    \langle W \rangle_{O} /k_B T=  \frac{\lambda_{f}}{2 \lambda_{i}}(1 + \tau_P c_{2}^*)^2 + \frac{(\tau_P c_{2}^*)^2 \gamma}{ \lambda_{i} \tau_P}  
     - \ln(1+c_{2}^*\tau_P)- \frac{1}{2}.
    \label{eq:average_work}
\end{equation}

Fig. \ref{fig:theoprot} displays representative optimal protocols for the transformation $\lambda_{i}\rightarrow \lambda_{f}$ at different parameters (solid lines). Notice that the optimal protocol requires discontinuities in the control of $\lambda$ at $t=0$ and $t=\tau_P$, which are a common feature of minimum-dissipation protocols~\cite{PhysRevE.104.L022101,Blaber_2023}. 

In this scenario, we experimentally explore and compare optimal and sub-optimal (linear) protocols by varying the protocol duration, $\tau_P$, and the modulation amplitude, $\Delta \lambda = \lambda_f - \lambda_i$. By controlling these parameters, we can systematically tune how far the process is driven out of equilibrium and investigate the associated energetic exchanges. 

To this end, we employ a system that allows trapping and tracking particles in a harmonic potential with high temporal resolution and fast modulation control, enabling the implementation of the sharp endpoint changes shown in Fig. \ref{fig:theoprot}. 

%##########################################################################

\section{Experimental setup}\label{sec:exp}

\begin{figure}
\centerline{\includegraphics[width=8cm]{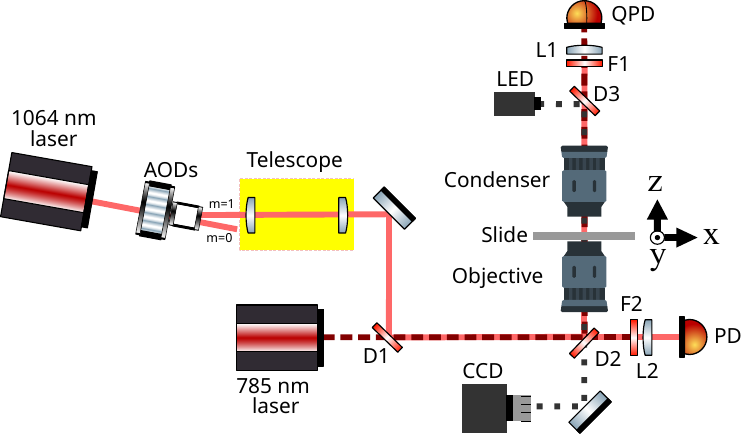}}
\caption{\textbf{Experimental setup.}
Schematic of the optical tweezers apparatus used in this work.
An infrared laser ($1064\,\mathrm{nm}$) creates the optical trap,
whose power is modulated by acousto-optic deflectors (AODs) and monitored by a photodetector (PD). 
The particle position is measured using a quadrant photodiode (QPD)
with a collinearly aligned probing beam ($785\,\mathrm{nm}$).
All labeled components are described in Sec.~\ref{sec:exp}.} 
\label{fig:setup}
\end{figure}

A sketch of the experimental apparatus used in this work is shown in Fig.~\ref{fig:setup}, and more details can be found in Refs. \cite{martins2024studies, martins2025brief}. 

In the trapping system, a Gaussian infrared laser beam ($1064\,\mathrm{nm}$) is focused through a $63\times$ ($\mathrm{NA}=1.32$) immersion objective, creating a three-dimensional harmonic potential that traps single $2\,\mathrm{\mu m}$ silica beads suspended in purified water at room temperature ($T \approx 293\,\mathrm{K}$). The beam is expanded using a telescope to fill the entrance pupil of the objective to maximize trapping efficiency. 

To perform the compressing protocols, it is necessary to control the trap stiffness of the potential, which is proportional to the beam intensity. For that, two perpendicular acousto-optic deflectors (AODs) are driven by an arbitrary waveform generator whose signal is first amplified by a homemade linear amplifier. By adjusting the amplitude of the radio-frequency signal driving one of the AODs, the intensity of the first-order diffracted beam can be precisely controlled. It is worth noting that the bandwidth of the arbitrary waveform generator sets the ultimate limit for the modulation speed; discontinuities in the signal amplitude introduce a time delay of approximately $100\,\mathrm{\mu s}$. 
A portion of the infrared light is diverted and focused onto a photodetector (PD) using lens L2 to monitor the laser intensity.

The particles in the sample are located and monitored using bright-field imaging with a CCD camera. However, for high-speed position detection, which is essential for resolving the particle’s dynamics at short time scales, we employ a forward-scattering interferometric technique. For that, we use a collimated $785\,\mathrm{nm}$ probe laser, aligned collinearly with the trapping beam. The resulting interference pattern between the scattered and transmitted light is collected by a $\mathrm{NA}=0.53$ condenser lens and directed onto a homemade quadrant photodiode (QPD), allowing us to measure the three-dimensional motion of the particle at a sampling frequency of $f_s = 100 \ \mathrm{kHz}$. A lens (L1) is used to focus the beam on the sensitive area of the QPD.

Control of the AODs, as well as acquisition of the PD and QPD signals, is performed using a high-speed multifunction data acquisition (DAQ) board. Mirrors and dichroic filters (D1, D2, and D3) direct the beams along the desired optical paths, while an interferometric filter (F1) prevents the trapping light from reaching the QPD, and a high-pass filter (F2) blocks the detection beam at the PD entrance.

%##########################################################################

\section{Protocols application and results}\label{sec:results}

The driving protocols are implemented by dynamically controlling the laser intensity while simultaneously recording the particle trajectory and the PD output signal, $V_{PD}$. Monitoring $V_{PD}$ is essential to synchronize the effective change in trap stiffness, which is affected by the finite response time of the optical modulation, with the measured particle dynamics.

In all theory-experiment comparisons below, the control parameter $\lambda(t)$ is taken directly from the experimentally realized stiffness protocol, reconstructed from the photodetector reading, rather than from the nominal waveform sent to the AODs. This distinction is particularly important for the shortest protocols, for which the finite modulation bandwidth smooths the endpoint jumps prescribed by the (ideal) analytical solution.

To ensure equilibrium initial conditions, the particle is allowed to thermalize in the potential at $\lambda_i = 1.5 \ pN/\mu m$ for a time $\tau_{eq} \approx 20 \ \mathrm{ms}$, longer than $\tau_{R} = \gamma / \lambda_i = 12 \ \mathrm{ms}$, the relaxation time. After this first equilibration step, the protocol is applied, either following a linear modulation of the trap stiffness or the optimal protocol curves discussed in Sec.~\ref{sec:theo}. At the end of the protocol, the system is allowed to equilibrate again with the thermal bath for a period longer than the relaxation time while kept in the final state, corresponding to $\lambda_f$. Note that right after finishing the protocol, the system is generally not in equilibrium, and the particle continues to exchange energy until full thermalization. The entire cycle is then repeated $N = 10^4$ times to build the ensemble statistics.

\begin{figure}
\centering
\includegraphics[width=8cm]{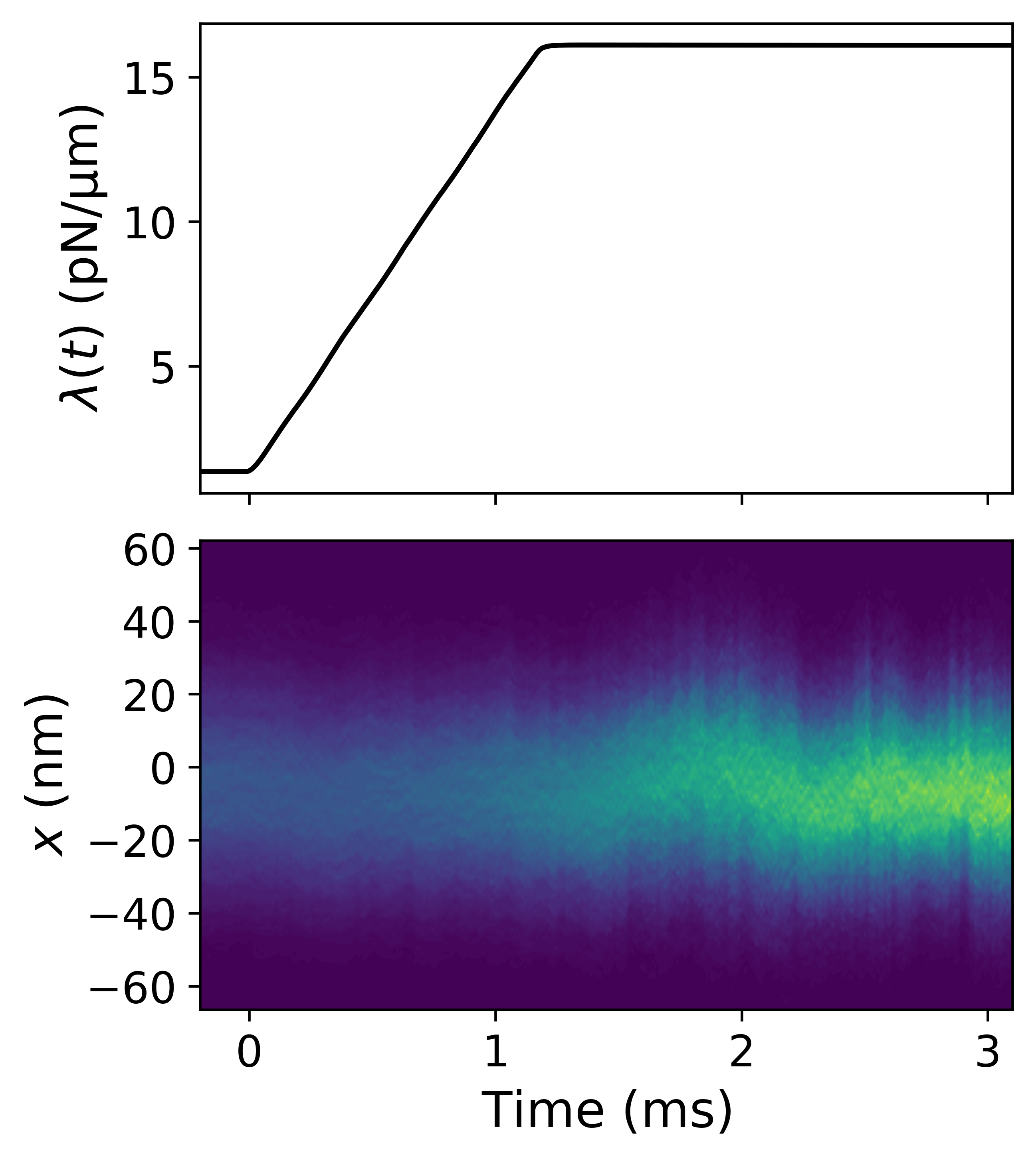}
\caption{\textbf{Compressing protocols}. Time evolutions of the trap stiffness (top) and position probability distribution (bottom) for a linear protocol with $\Delta \lambda= 10 \lambda_i$ and $\tau_P = 0.1\tau_R$. The number of trajectories is $N=10^4$, and the number of bins in each histogram is $100$.}
\label{fig:xandlambda}
\end{figure}

Figure \ref{fig:xandlambda} illustrates the data recorded for a linear compression protocol for a set of $10^4$ trajectories. It presents the time evolution of the trap stiffness (top) and the resulting position probability distribution (bottom). The latter was obtained by constructing instantaneous histograms from the ensemble of positions at each time. Initially, the particle has a larger width of the distribution since it is trapped in a less confining state. After applying the protocol, the width is narrowed, constraining the particle's motion around the center of the trap, $x=0$. 

From the values of the particle's trajectory, $ x_{j}$\footnote{The same procedure can be applied to the $y$ and $z$ directions, but throughout this work, the focus will be on the one-dimensional behavior in $x$.}, and the finite difference of the trap stiffness, $\lambda_{j+1} - \lambda_{j}$\footnote{Note that $x_{j}=x(t_j)$, and $\lambda_{j}=\lambda(t_j)$, with $t_j = j \Delta t$, $\Delta t=1/f_s$, and $n=f_s \tau_P$. The sum in Eq.~\ref{eq:workfinal} runs over $j=0,\ldots,n-1$. The required calibration procedures are described in Appendix~\ref{sec:kappaandS}.}, it is possible to compute the discrete work along the trajectory (Eq. \ref{eq:work}) \footnote{The optimal protocols present sudden jumps. Such discontinuities contribute as follows:
$$
    W_{jump} = \frac{\lambda_+ x_+^2}{2} - \frac{\lambda_- x_-^2}{2},
$$
where the subscript '$-$' represents the point just before the jump, and '$+$' represents the point just after it. However, experimentally, the optical power is varied continuously, and when the acquisition frequency is sufficiently high, $x_+ \approx x_-$, and Eq.~\ref{eq:workfinal} remains valid.}:
\begin{equation}
    W_n = \frac{1}{2}\Delta t \sum_{j=0}^{n-1} \frac{\lambda_{j+1} - \lambda_{j}}{\Delta t} x_j^2 = \frac{1}{2} \sum_{j=0}^{n-1} (\lambda_{j+1} - \lambda_{j}) x_j^2.
    \label{eq:workfinal}
\end{equation}

\begin{figure}
\centering
\includegraphics[width=8cm]{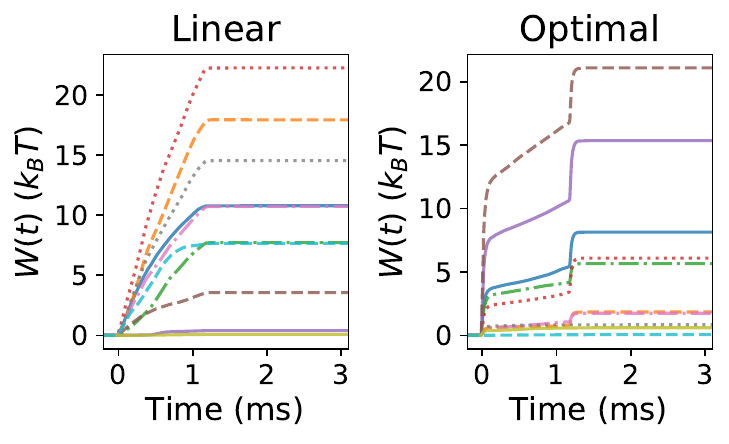}
\caption{\textbf{Time evolution of the cumulative work}. Results for $10$ realizations of linear (left) and optimal (right) protocols. For both cases, $\Delta \lambda=10 \lambda_i$, and $\tau_P=0.1 \tau_R$.}
\label{fig:Workloop}
\end{figure}

Figure~\ref{fig:Workloop} shows the time evolution of the cumulative work during the execution of a linear (left) and an optimal (right) protocol, for $10$ individual trajectories. In both cases, the work evolves differently in time for each realization, reflecting the intrinsically stochastic nature of the particle dynamics.
Moreover, according to Eq. \ref{eq:workfinal}, the work is accumulated exclusively during the application of the protocol. After $t>\tau_P$, the trap stiffness is kept constant, and no further work is performed, whereas heat and internal energy can still evolve during the relaxation process.

For the optimal protocol, the smoother accumulation observed reflects the finite control bandwidth discussed above. Since we use the experimentally implemented stiffness, the finite-time response effect is already included in the analysis.

\begin{figure}
\centering
\includegraphics[width=8cm]{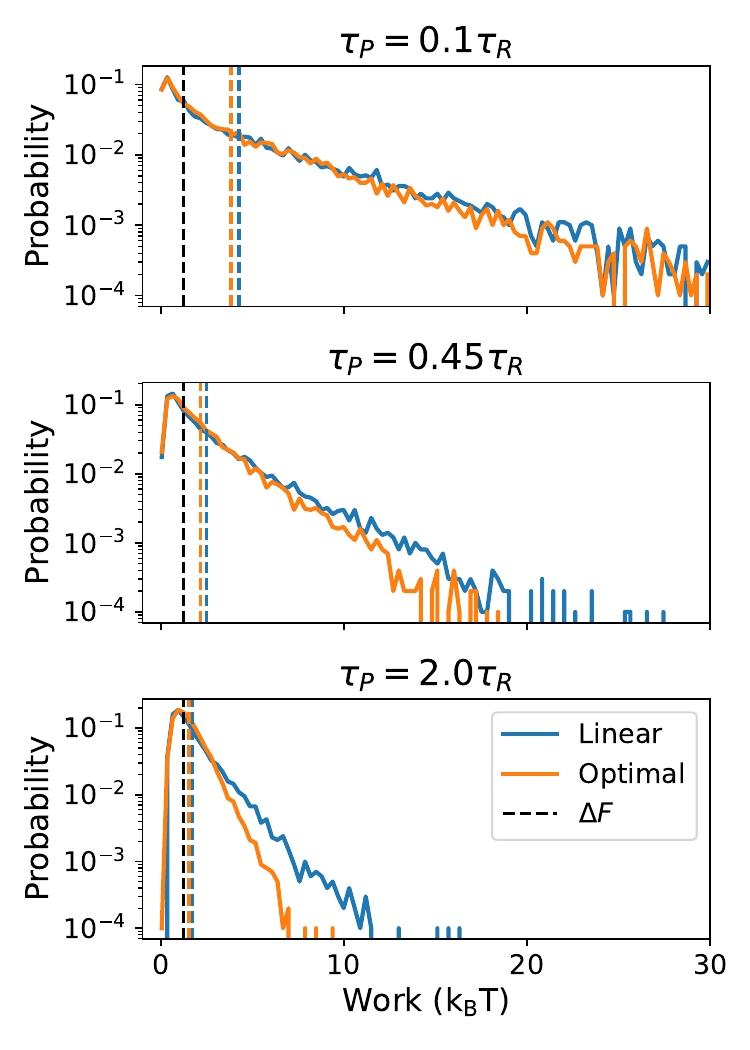}
\vspace{-3mm}
\caption{\textbf{Work probability distributions} for $\Delta \lambda= 10 \lambda_i$ and different protocol times: $\tau_P = 0.1\tau_R$ (top), $\tau_P = 0.45\tau_R$ (middle), and $\tau_P = 2\tau_R$ (bottom). Data were obtained from $N=10^4$ trajectories for linear protocols (blue) and optimal ones (orange). The black dashed line represents the free energy difference, and the colored dashed lines indicate the work average values. The number of bins is $100$.}
\label{fig:Whisto_2}
\end{figure}

Computing the cumulative work in each cycle, $W(\tau_P)$, we examine the resulting probability distributions across the ensemble of $N=10^4$ trajectories, as shown in Fig.~\ref{fig:Whisto_2}. Here, we explore both linear and optimal protocols with varying protocol times and fixed modulation amplitude, $\Delta \lambda= 10 \lambda_i$.

Clearly, the distributions of linear and optimal protocols are distinct, particularly in the tails, with the average value of the optimal case being consistently lower than that of the linear. Nevertheless, the mean work for both protocols converges to $\Delta F$ for longer protocol times, as expected from the Clausius inequality, with equality reached in the quasi-static limit. The width of the distributions also decreases with increasing protocol duration, ideally approaching a delta function centered at $W = \Delta F$ in the quasi-static limit.
Although the mean work is always greater than the free energy difference, a considerable number of events show lower work. This clearly does not violate the second law of thermodynamics, as it is based on averages rather than individual events.

\begin{figure}
    \centering
    \includegraphics[width=7.6cm]{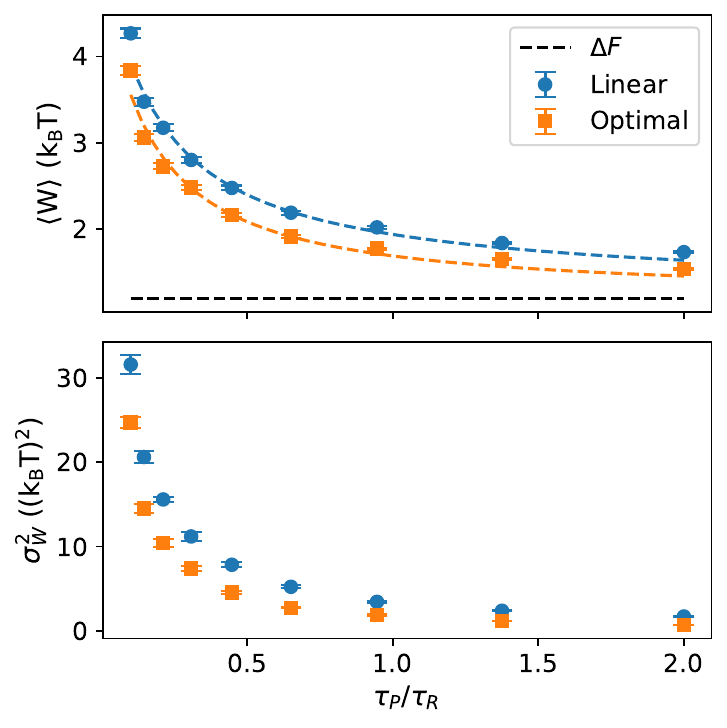}  
    \vspace{-3mm}
    \caption{\textbf{Work statistics for varying protocol durations}. Curves for average work (top) and work variance (bottom), for an ensemble of $10^4$ trajectories with fixed $\Delta \lambda = 10 \lambda_i$. Experimental results for linear (blue) and optimal (orange) protocols are represented by dots and theoretical curves as dashed lines. The free energy difference is represented by a black dashed line.}
    \label{fig:Resultstimes}
\end{figure}

To quantify this behavior more systematically, Figure~\ref{fig:Resultstimes} presents the statistical results for both types of protocols, showing the mean and variance, $\sigma_W^2 = \langle (W - \langle W \rangle)^2 \rangle$, of work across a wide range of protocol durations\footnote{The uncertainty of the mean work is the standard deviation of the work distribution, $\sigma_W$, divided by the square root of the number of trajectories, $\delta \langle W \rangle = \sigma_W/\sqrt{N}$, where $N=10^4$.
For the work variance, the statistical uncertainty is estimated using a block-averaging procedure. 
The full dataset is divided into $n_{\mathrm{b}}=10$ independent blocks, each containing 1000 trajectories. 
For each block $i$, we compute the variance $\sigma^2_{W,i}$ of the work distribution.
The uncertainty of the variance is then estimated as
$$
\delta\left[\sigma_W^2\right]
=
\frac{\mathrm{std}\left(\{\sigma^2_{W,i}\}_{i=1}^{n_{\mathrm{b}}}\right)}{\sqrt{n_{\mathrm{b}}}} .
$$}. 

For all cases, we consistently observe that both the average work and the variance for the optimal protocols are lower than those of the linear ones. Nevertheless, the difference between the two protocols is expected to vanish for very short or very long protocol durations. When the process is performed very quickly, the average work converges to that of a step process (Eq.~\ref{eq:wstep}). Conversely, for very long $\tau_P$, the work approaches the free energy difference, $\Delta F$ (Eq. \ref{eq:deltaftheo}), for both protocols. This indicates that the largest benefits of optimal driving are expected at intermediate, short-but-finite durations, before reaching the step limit. Furthermore, the average work shows excellent agreement with the theoretical prediction. For the work variance, as also seen in Fig.~\ref{fig:Whisto_2}, it decreases and approaches zero for long protocol durations, as expected.

Beyond the reduction of the mean dissipated work, the visible differences in the distribution tails suggest that, in our system, optimal driving also suppresses rare, highly dissipative events, a feature relevant to the full characterization of nonequilibrium work statistics.
This is clearly visible in Fig.~\ref{fig:Whisto_2} as a systematic depletion of the large-W tail for the optimal protocol. The broader interpretation of this simultaneous reduction of mean work, variance, and high-work events is discussed in Sec.~\ref{sec:discussion}.

\begin{figure}
    \centering
    \includegraphics[width=7.6cm]{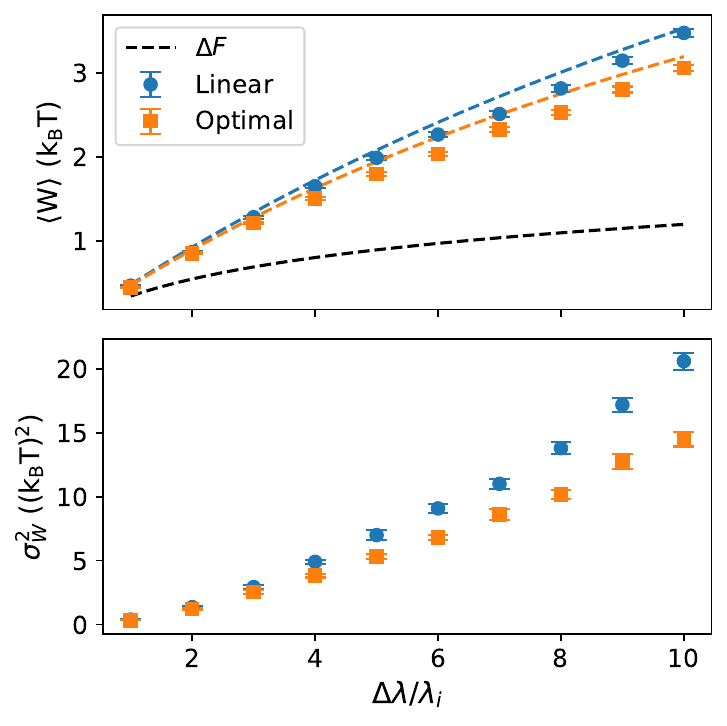}
    \vspace{-3mm}
    \caption{\textbf{Work statistics for varying modulation amplitudes}. Curves for average work (top) and work variance (bottom), for an ensemble of $10^4$ trajectories with fixed $\tau_P = 0.145 \tau_R$ and modulation amplitude from $\Delta \lambda = 1\,\lambda_i$ to $10\,\lambda_i$. Experimental results for linear (blue) and optimal (orange) protocols are represented by dots and theoretical curves as dashed lines. The free energy difference is represented by a black dashed line.}
    \label{fig:Resultsamps}
\end{figure}

Figure~\ref{fig:Resultsamps} presents the experimental results for a fixed protocol duration, $\tau_P = 0.145\,\tau_R$, while varying the modulation amplitude. As the amplitude increases, the system is driven further from equilibrium, and both the mean work and variance grow. The optimal protocol consistently exhibits lower values than the linear one, and the difference becomes increasingly significant for larger $\Delta \lambda$, 
showing that the benefit of optimized driving increases as the system is driven farther from equilibrium within the investigated parameter range.

%#######################################################################

\section{Discussion}\label{sec:discussion}

Using an optical-tweezers platform, we have implemented and benchmarked the finite-time minimum-work compression protocol analytically derived by Schmiedl and Seifert for an overdamped Brownian particle in a harmonic trap~\cite{schmiedl2007optimal}. By comparing it directly with sub-optimal linear driving across a broad range of durations and amplitudes, we provide a direct experimental validation of an exact analytical minimum-work solution in stochastic thermodynamics and reveal additional
features of its work statistics.

Across the investigated parameter range, the optimal protocol consistently yields lower mean work and work variance than linear driving, with the largest differences occurring at short-but-finite durations ($\tau_P \lesssim \tau_R$, i.e., before the instantaneous step limit) and larger compression amplitudes. 

A key experimental insight of this work is the robustness of the protocol's advantage under realistic instrumental constraints. Although the ideal protocol prescribes discontinuous jumps in trap stiffness at the endpoints, our implementation is limited by the finite modulation bandwidth of the arbitrary waveform generator. Even with the smoothing of these jumps, the reduction in thermodynamic cost remains clearly observable. 
This shows that the advantage is experimentally accessible despite the finite response time of the present implementation, a feature potentially relevant to other
bandwidth-limited control platforms.

The reduction in work fluctuations also has practical implications. Broad and non-Gaussian work distributions can affect the finite-sample properties of nonequilibrium free-energy estimators \cite{jarzynski1997nonequilibrium,PhysRevLett.100.190601,PhysRevE.103.032146}. In the range investigated here, the minimum-mean-work protocol reduces both the work variance and the high-work tail relative to linear driving, even under bandwidth-limited actuation. Although these quantities were not introduced as separate optimization objectives, this result is consistent with the growing interest in higher moments as relevant control targets in strongly fluctuating systems~\cite{Blaber_2023}. In the slow, near-equilibrium Gaussian regime discussed in Sec.~\ref{sec:intro}, reductions in dissipation and variance are directly related. Outside this regime, however, minimizing the mean work does not generally imply minimizing the variance or higher cumulants. We therefore interpret the simultaneous improvement as a noteworthy experimental property of the harmonic-compression setting studied here, while its broader validity remains an open theoretical question.

In contrast to shortcut-to-isothermality protocols, whose goal is to enforce equilibration through auxiliary control, and to optimal-transport experiments relying on arbitrary potential shaping, here the control acts through a single physical parameter -- the trap stiffness -- allowing a direct comparison between measurement and the analytical solution in~\cite{schmiedl2007optimal}.

In summary, this study establishes a reference point for investigating more complex control strategies. While we focused on single-parameter harmonic control, the precision of the current setup opens the way for future explorations of multi-parameter driving, active matter systems, and non-trivial potentials where analytical solutions are unavailable~\cite{Deffner_2020, Blaber_2023, bonancca2014optimal, PhysRevE.94.052106, bonancca2018minimal, kamizaki2022performance}.

\textbf{Data availability:} All the pre-processed datasets and analysis code presented are openly available on the Zenodo server at DOI:10.5281/zenodo.19705797 \cite{ZenodoDOI}.

%#######################################################################

\section*{Acknowledgment}

The authors acknowledge the financial support from the following research agencies: CAPES (process 88887.370240/2019-00), CNPq (process 131013/2020-3), and FAPESP (grant 2019/27471-0). A.H.A.M. acknowledges support from the National Science Centre, Poland (SONATINA-9, grant agreement no. UMO-2025/56/C/ST2/00368 entitled “Modularne kwantowe urządzenia termiczne: integrowanie termicznych funkcjonalności”).

%#######################################################################

\appendix

\section{Characterization of the optical potential}
\label{sec:kappaandS}

In this work, we use the Power Spectrum Density (PSD) analysis \cite{Berg-Sorensen2004} to determine the trap stiffness of the potential, $\kappa$, and the amplification factor, $S$, responsible for converting the QPD output, $x_{QPD}$, from volts to meters by using $x = x_{QPD}/S$. 

The basic procedure is to trap one particle in the static potential and track its position (in volts), $x_{QPD, j}$,  at a fixed sampling frequency $f_s = 100 \ \mathrm{kHz}$ for a total time $T_s=30 \ \mathrm{s}$. 

The PSD is computed directly from the measured trajectories using a fast Fourier transform within Welch’s method \cite{welch1967use}. The number of points used for the fast Fourier transform is $N_{FFT} = 8 \times 10^{5}$, and the number of points to overlap between segments is $0.9 \ N_{FFT}$. For each power level, ten independent measurements were performed, and the final PSD corresponds to the average of these measurements.
 
The PSD is given by \cite{Berg-Sorensen2004, jones2015optical}:
\begin{equation}
   P(f) = \frac{S^2 D/\pi^2}{f_c^2+f^2},
    \label{eq:normalPSD}
\end{equation}
being $D = k_BT/\gamma$ the diffusion coefficient, $S$ the amplification factor, and $f_c=\kappa (2 \pi \gamma)^{-1}$ the corner frequency, which depends on the trap stiffness and the particle friction coefficient. 

\begin{figure}[tb]
\vspace{2mm}
\centerline{\includegraphics[width=8.5cm]{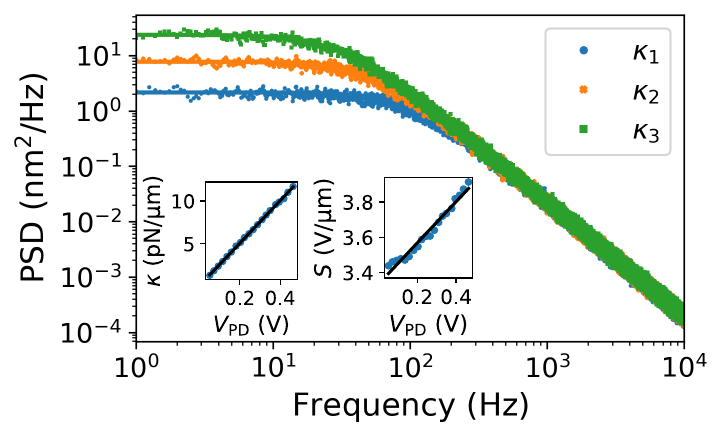}}
\vspace{-4mm}
\caption{Power spectrum density analysis corresponding to three different laser powers, from the lowest (green) to the highest (blue) stiffness, where ($\kappa_{1}>\kappa_{2}>\kappa_{3}$). Inset: calibration curves for trap stiffness (left) and amplification factor (right). Linear fits are represented as black solid lines.}
\label{fig:ts1}
\end{figure}

Figure \ref{fig:ts1} shows the curves obtained for three different trapping beam powers and their respective fits using Eq. \ref{eq:normalPSD}. It is important to note that precise control of the beam intensity is achieved through prior calibration of the AOD response. More details can be found in Refs. \cite{martins2024studies, martins2025brief}. 

The inset of Fig. \ref{fig:ts1} presents the trap stiffness (left) and amplification factor (right) obtained from the PSD fits. As expected, the trap stiffness is linearly dependent on the laser intensity (proportional to the mean PD output voltage, $\langle V_{PD} \rangle$).
However, although the amplification factor should remain constant in an ideal system, it also varies. This happens because the change in the trapping laser's intensity shifts the equilibrium position along the $z$-axis, influencing the sensitivity of the QPD and, consequently, the amplification factor. 
In this way, linear fits, shown in Fig.~\ref{fig:ts1}, are used to estimate the trap stiffness and the amplification factor for a given laser power, characterized by the mean photodetector output voltage, $\langle V_{PD} \rangle$.

More specifically, to determine the amplification factor used throughout the compression protocols, we consider the average PD output signal measured during the equilibrium period in the initial state. This approach is adopted because modeling the time evolution of the particle trajectory along the $z$-axis during the protocol is challenging. In general, for short protocol times and modulation amplitudes, the particle is expected to move less in the $z$-axis, such that it remains closer to the initial position, and this calibration provides a good approximation. However, the change is more considerable for longer protocols and larger changes in the trap stiffness.

To assess the impact of axial (z-axis) shifts on our work measurements, we recalculated the work values using the amplification factor obtained from the average PD output signal measured in the final equilibrium state as a limiting case. This provides a conservative estimate of the largest calibration-induced deviation. For the largest change in trap power ($\Delta \lambda = 10\lambda_i$), the deviation relative to the results presented in this work is approximately $20\%$. This estimate should therefore be interpreted as a conservative consistency check for the worst-case scenario, rather than a true uncertainty. Moreover, since the same calibration procedure is consistently applied to both optimal and linear protocols, the relative comparison between them remains essentially unchanged, thereby preserving the main conclusions of this work.

%##########################################################################

\bibliography{refs}

%##########################################################################

\end{document}